\documentclass[preprint]{aastex}
\usepackage{emulateapj5}
\usepackage{palatcm,eucal}

\newcommand{\unitspace}{\ensuremath{\;}}
\newcommand{\usp}{\unitspace}
\newcommand{\unitstyle}[1]{\ensuremath{\mathrm{#1}}}
\newcommand{\power}[2]{\ensuremath{{#1}^{#2}}}

\newcommand{\kilo}{\unitstyle{k}}
\newcommand{\Mega}{\unitstyle{M}}

\newcommand{\cm}{\unitstyle{cm}}
\newcommand{\gram}{\unitstyle{g}}

\newcommand{\meter}{\unitstyle{m}}

\newcommand{\second}{\unitstyle{s}}

\newcommand{\K}{\unitstyle{K}}	

\newcommand{\grampercc}{\gram\usp\power{\cm}{-3}} 
\newcommand{\grampersquarecm}{\gram\usp\power{\cm}{-2}} 
\newcommand{\GramPerCc}{\grampercc}
\newcommand{\GramPerSc}{\grampersquarecm}
\newcommand{\erg}{\unitstyle{ergs}} 

\newcommand{\ergspersecond}{\erg\unitspace\power{\second}{-1}}

\newcommand{\eV}{\unitstyle{eV}}	
\newcommand{\MeV}{\Mega\eV} 

\newcommand{\Msun}{\ensuremath{M_\odot}}

\newcommand{\hour}{\unitstyle{hr}} 
\newcommand{\yr}{\unitstyle{yr}}	
\newcommand{\km}{\kilo\meter}	

\newcommand{\unit}[2]{\ensuremath{#1\usp\mathrm{#2}}}

\newcommand{\kB}{\ensuremath{k_\mathrm{B}}} 
\newcommand{\pF}{\ensuremath{p_\mathrm{F}}} 
\newcommand{\EF}{\ensuremath{\varepsilon_\mathrm{F}}} 
\newcommand{\mb}{\ensuremath{m_\mathrm{u}}} 
\newcommand{\me}{\ensuremath{m_\mathrm{e}}} 
\newcommand{\mpr}{\ensuremath{m_\mathrm{p}}} 
\newcommand{\mn}{\ensuremath{m_\mathrm{n}}} 

\newcommand{\dif}{\ensuremath{\mathrm{d}}}

\newcommand{\ee}[1]{\ensuremath{\times 10^{#1}}}

\newcommand{\satellite}[1]{\emph{#1}}

\newcommand{\asca}{\satellite{ASCA}}

\newcommand{\chandra}{\satellite{Chandra}}
\newcommand{\einstein}{\satellite{EINSTEIN}}
\newcommand{\exosat}{\satellite{EXOSAT}}
\newcommand{\rosat}{\satellite{ROSAT}}

\newcommand{\xmm}{\satellite{XMM}}

\RequirePackage{amssymb}
\RequirePackage{amsbsy}
\newcommand{\bvec}[1]{\ensuremath{\boldsymbol{#1}}} 
\newcommand{\grad}{\bvec{\nabla}} 

\newcommand{\nuclei}[2]{\ensuremath{\mathrm{^{#1}#2}}}
\newcommand{\Hyd}{\ensuremath{\mathrm{H}}}
\newcommand{\He}{\nuclei{4}{He}}
\newcommand{\C}{\nuclei{12}{C}}
\newcommand{\Fe}{\nuclei{56}{Fe}}
\newcommand{\Ru}{\nuclei{104}{Ru}}
\newcommand{\sigmaSB}{\ensuremath{\sigma_\mathrm{R}}}
\newcommand{\Te}{\ensuremath{T_\mathrm{eff}}}
\newcommand{\Teo}{\ensuremath{T_\mathrm{eff,\infty}}}
\newcommand{\Tb}{\ensuremath{T_\mathrm{b}}}
\newcommand{\yb}{\ensuremath{y_\mathrm{b}}}
\newcommand{\yi}{\ensuremath{y_\mathrm{i}}}
\newcommand{\opac}{\ensuremath{\kappa}}
\newcommand{\opacrad}{\ensuremath{\opac_\mathrm{r}}}
\newcommand{\opaccond}{\ensuremath{\opac_\mathrm{c}}}
\newcommand{\Zbar}{\ensuremath{\langle Z\rangle}}
\newcommand{\ZZbar}{\ensuremath{\langle Z^2\rangle}}
\newcommand{\Abar}{\ensuremath{\langle A\rangle}}
\newcommand{\Lej}{\ensuremath{\Lambda_{\mathrm{e},j}}}
\newcommand{\Leimp}{\ensuremath{\Lambda_\mathrm{e,imp}}}
\newcommand{\source}[3]{#1~#2$#3$} 
\newcommand{\aql}{Aql~X-1}
\newcommand{\cen}{Cen~X-4}
\newcommand{\sax}{\source{SAX}{J1808.4}{-3658}}
\newcommand{\D}{\ensuremath{\mathcal{D}}}
\newcommand{\vdr}{\ensuremath{\bvec{w}}}
\newcommand{\taudr}{\ensuremath{\tau_\mathrm{dr}}}
\newcommand{\Hp}{\ensuremath{H_p}}
\newcommand{\Htr}{\ensuremath{H_2}}
\newcommand{\yss}{\ensuremath{y_\mathrm{ss}}}
\newcommand{\GammaM}{\ensuremath{\Gamma_\mathrm{M}}}
\newcommand{\Ye}{\ensuremath{Y_\mathrm{e}}}
\newcommand{\nel}{\ensuremath{n_\mathrm{e}}}

\begin{document}


\slugcomment{Final version 3 April 2002}
\journalinfo{To appear in \textsc{The Astrophysical Journal}}

\title{%
  Variability in the Thermal Emission from Accreting Neutron Star
  Transients
}

\author{%
  Edward F. Brown
}
\affil{%
  University of Chicago, Enrico Fermi Institute, 
  5640 S. Ellis Ave, Chicago, IL 60637}
  \email{brown@flash.uchicago.edu}
\author{%
  Lars Bildsten
}
\affil{%
  Institute for Theoretical Physics and Department of Physics, Kohn Hall,
  University of California, Santa Barbara, CA 93106}
  \email{bildsten@itp.ucsb.edu}
\and
\author{%
  Philip Chang
}
\affil{
  Department of Physics, Broida Hall, University of
  California, Santa Barbara, CA 93106}
  \email{pchang@physics.ucsb.edu}

\shorttitle{Variability in Neutron Star Transients}
\shortauthors{Brown, Bildsten, \& Chang}


\begin{abstract}

    The composition of the outer 100\usp\meter\ of a neutron star sets
    the heat flux that flows outwards from the core.  For an accreting
    neutron star in an X-ray transient, the \emph{thermal} quiescent
    flux depends sensitively on the amount of hydrogen and helium
    remaining on the surface after an accretion outburst and on the
    composition of the underlying ashes of previous \Hyd/\He\ burning.
    Because \Hyd/\He\ has a higher thermal conductivity, a larger mass
    of \Hyd/\He\ implies a shallower thermal gradient through the low
    density envelope and hence a higher effective temperature for a
    given core temperature.  The mass of residual \Hyd\ and \He\ varies
    from outburst to outburst, so the thermal quiescent flux is variable
    even though the core temperature is constant for timescales
    $\lesssim 10^{4}\usp\yr$.  Heavy elements settle from a \Hyd/\He\
    envelope in a few hours; we therefore model the quiescent envelope
    as two distinct layers, \Hyd/\He\ over heavier elements, and treat
    the mass of \Hyd/\He\ as a free parameter.  We find that the
    emergent thermal quiescent flux can vary by a factor of 2 to 3
    between different quiescent epochs.  The variation is more
    pronounced at lower interior temperatures, making systems with low
    quiescent luminosities and frequent outbursts, such as \sax, ideal
    candidates from which to observe this effect.  Because the ashes of
    \Hyd/\He\ burning are heavier than \Fe, their thermal conductivity
    is greatly reduced.  This increases the inferred crust temperature
    beyond previous estimates for a given effective temperature.  We
    survey this effect for different ash compositions and apply our
    calculations to \cen, \aql, and \sax.  In the case of \aql, the
    inferred high interior temperature suggests that neutrino cooling
    contributes to the neutron star's thermal balance.

\end{abstract}



\section{INTRODUCTION}\label{sec:introduction}

The orbiting X-ray observatories \chandra\ and \xmm\ have dramatically
improved our understanding of soft X-ray transients (SXTs): binaries
containing a neutron star or black hole primary and having well-defined
accretion outbursts separated by long periods of quiescence.  These
objects are typically defined as having a ratio of outburst flux to
quiescent flux $> 1000$.  Two puzzles are pertinent to this work.  The
first is whether the thermal component of the neutron star's quiescent
luminosity is powered by accretion or by thermal emission from the
cooling core.  This has implications for the observed contrast in the
quiescent luminosity between transient black holes and neutron stars
\citep{narayan97a,menou99,garcia.mcclintock.ea:new,kong.ea:bh_xrn}.

\citet{brown98:transients} showed that compression-induced
reactions---electron captures, neutron emissions, and pycnonuclear
reactions \citep{bisnovatyi-kogan79:_noneq_x,sato79,haensel90a}---in
the inner crust of an accreting neutron star release enough heat to
power a cooling luminosity of order $10^{33}\usp\ergspersecond$ in
quiescence.  In the absence of neutrino emission from the core the
quiescent thermal flux is proportional to the mean outburst flux
\citep{brown98:transients,colpi.geppert.ea:charting}.  Motivated by
the match between the expected quiescent luminosity and that observed
from neutron star SXT's in quiescence,
\citet{rutledge.ea.99:refit,rutledge.ea.00:BH} fit archival \rosat\
and \asca\ observations of \aql, \cen, \source{4U}{1608}{-522}, and
\source{4U}{2129}{+47} with realistic \Hyd\ (or \He) atmosphere
spectra and found that the emission could be explained as thermal
emission from an area of radius $\approx 10\usp\km$.  Further
\chandra\ observations of \cen\ \citep{rutledge.ea.01a:cen}, \aql\
\citep{rutledge.ea.01b:aqlx1}, and \source{KS}{1731}{-260}
\citep{wijnands:ks1731,rutledge.ea.01:ks1731}, as well as quiescent
transient neutron star identifications in $\omega$~Cen
\citep{rutledge.ea:NGC5139}, NGC~6440 \citep{pooley.ea:NGC6440}, and
47~Tuc X5 and X7 \citep{heinke02:47Tuc-Lq} confirm that the quiescent
spectra of neutron star transients is consistent with being thermal
emission (effective temperature $\kB\Te\gtrsim 100\usp\eV$) from a
\Hyd\ photosphere, plus, in most cases, an additional hard power-law
component.  The origin of the power-law tail remains uncertain (for a
review of proposed mechanisms, see \citealt{campana98b}; also see
\citealt{menou.mcclintock:quiescent-emission}).

The second puzzle is the source of the observed variability in the
quiescent emission on timescales $> \unit{1}{d}$.  Indeed, it was in
part because of an apparent \emph{increase} in the quiescent intensity
of \cen\ between 1980 (\einstein) and 1984 (\exosat) that led
\citet{vanparadijs87} to discount the possibility that the observed
emission was intrinsic to the neutron star, i.e., not powered by
accretion.  \rosat/HRI observations of \cen\ \citep{campana97} revealed
that the intensity decreased by a factor of $\approx 3$ over 4 days.
Similarly, there was a fractional decrease of 40\usp\% in the observed
intensity between an \asca\ observation and one 5\usp\yr\ later with
\chandra, although this could be attributed to variability in the
power-law component only \citep{rutledge.ea.01a:cen}.  A comparison of
\aql\ observations taken with \chandra/ACIS-S, \rosat/PSPC, and \asca\
\citep{rutledge.ea.01b:aqlx1} indicates variability by a factor of 2
over a timescale of roughly 8\usp\yr.  Note that in this case there were
several intervening outbursts between the different observations.  In
both cases, there was no short-timescale ($\lesssim 10^{4}\usp\second$)
variability detected \citep{rutledge.ea.01a:cen,rutledge.ea.01b:aqlx1}.
More recently, \citet{rutledge.bildsten.ea:intensity} used the
\chandra/ACIS-S to take four ``snapshots'' of \aql\ after a recent
outburst.  The intensity was observed to decrease by a factor $\approx
0.5$ over 3 months and then increase by a factor $\approx 1.4$ over 1
month.  In addition, short-timescale variability was found in the last
observation.

The standard interpretation is that the observed variability is caused
by fluctuations in the quiescent accretion rate
\citep{vanparadijs87,campana97,brown98:transients,rutledge.ea.01a:cen,
rutledge.ea.01b:aqlx1,menou.mcclintock:quiescent-emission,dubus.ea:DIM-SXT}.
 In this manuscript, we describe a previously overlooked cause of
variability in the intrinsic quiescent thermal emission: a changing
envelope\footnote{In this paper ``envelope'' means the outermost layer
of the neutron star where the thermal gradient is significant.  This
is distinct from the photosphere, where the emergent continuum
spectrum forms.} composition.  Even if accretion completely halts in
quiescence, the neutron star's envelope will have a different
stratification following each outburst.  This varying composition can
change the quiescent flux by a factor of 2 to 3 for a fixed crust/core
temperature.

Previous calculations of the thermal structure of a cooling unmagnetized
neutron star considered the difference between a purely \Fe\ envelope
\citep{gudmundsson83} and one composed of light elements (\Hyd, \He, and
\C) overlying \Fe\ \citep{potekhin97}.  Consider two hypothetical
neutron stars each with a core/crust temperature of $\Tb = 10^{8}\usp\K$
(typical of neutron star SXT's; see below); one star has a pure \Fe\
envelope and the other has \Hyd\ and \He\ at densities $<
10^{5}\usp\grampercc$.  As noticed by \citet{potekhin97}, the large
difference in opacity between a pure iron and a light element envelope
means that the effective temperature, $\Te$, for the light element
envelope is a factor of 1.6 hotter: $\Te(\Fe) = 1.1\ee{6}\usp\K$ and
$\Te(\mbox{light element}) = 1.8\ee{6}\usp\K$.  \citet{gudmundsson83}
first noticed that the thermal stratification is sensitive to the
opacity in the region where the electrons are semi-degenerate.
Coincidentally, it is in this region that the accreted \Hyd\ and \He\
unstably ignite.  As a result the thermal gradient through the envelope
depends on the mass of \Hyd\ and \He\ remaining after the
\emph{previous} accretion outburst.  Our calculation thus addresses
variations in the quiescent flux from one quiescent epoch to the next,
as the intervening outburst changes the composition and mass of the
outermost layers of the neutron star.  Changes in the quiescent
intensity over a timescale of months can also occur from differential
sedimentation of ions and residual accretion.  Neither of these
scenarios can explain short timescale ($<10^4\usp\second$) variability
such as just observed from \aql\ \citep{rutledge.bildsten.ea:intensity}.

This paper first (\S~\ref{sec:how-it-works}) describes in qualitative
terms the overall stratification of a quiescent neutron star transient.
Section~\ref{sec:microphysics} contains a summary of the relevant
microphysics in the calculation: the equation of state (EOS), diffusive
sedimentation of ions, and thermal transport.  In \S~\ref{sec:example}
we describe how changing the composition and stratification of the
envelope produces variations in the surface effective temperature.  This
calculation is then applied, in \S~\ref{sec:variation}, to \aql, \cen,
and \sax.  Implications and directions for future study are discussed in
\S~\ref{sec:summary}.

\section{THE COMPOSITION AND STRATIFICATION OF QUIESCENT NEUTRON STAR
ENVELOPES}
\label{sec:how-it-works}

In the absence of accretion, the thermal structure of the envelope is
determined, over durations much less than the cooling timescale of the
core (i.e., $\lesssim 10^{4}\usp\yr$), by the flux equation
\begin{equation}
    \frac{\dif}{\dif y} \left(\frac{T}{\Te}\right)^{4} = \frac{3}{4}
    \opac.  
    \label{eq:thermal}
\end{equation}
Here $\opac = (\opacrad^{-1} + \opaccond^{-1})^{-1}$ is the reciprocal
sum of the radiative opacity $\opacrad$ and the conductive opacity
$\opaccond$,
\begin{equation}
    \opaccond = \frac{16\sigmaSB T^{3}}{3\rho K},
    \label{eq:sigma}
\end{equation}
with $K$ and $\rho$ being the electron thermal conductivity and mass
density.  The spatial coordinate is just the column depth, $y =
\int_{r}^{\infty} \rho\,\dif r = p/g$ by hydrostatic balance.  We use
the Newtonian form of the thermal diffusion equation: the thickness of
the envelope is much less than the stellar radius, so that the
gravitational redshift $1+z\approx [1-2GM/(Rc^{2})]^{-1/2}$ is nearly
constant across the envelope and factors from
equation~(\ref{eq:thermal}).  All quantities in this manuscript refer to
proper quantities; in particular the effective temperature as observed
far away from the star is $\Teo = \Te (1+z)^{-1}$.

For a fixed envelope stratification, the flux equation
(\ref{eq:thermal}) guarantees a one-to-one mapping between $\Tb$ and
$\Te$.  The core temperature cannot change on timescales $\ll
10^{4}\usp\yr$, so if the envelope composition were constant, then the
basal effective temperature and luminosity would be unchanging from
quiescent epoch to quiescent epoch.  For an accreting neutron star the
envelope composition and stratification are not, however, fixed.  During
each outburst, \Hyd\ and \He\ are deposited onto the surface of the
neutron star.  After accumulation of a critical column
$y_\mathrm{ign}\sim 10^{8}\usp\GramPerSc$, the \Hyd\ and \He\ burn to
heavier elements (``ashes''), and the process then repeats.  As the
outburst wanes, there is a last episode of unstable burning (a type I
X-ray burst).  Accretion after this last type I burst deposits a
residual \Hyd/\He\ layer of column $\yi < y_\mathrm{ign}$ onto the ashes
of previous episodes of \Hyd/\He\ burning.  The depth of the light
element layer is effectively unconstrained, and as a result \Te\ can
vary even though \Tb\ is fixed.

The composition of the ashes depends on the nature of the \Hyd/\He\
burning \citep[for a recent review, see][]{bildsten.theory}; for most
accretion rates, the \He\ unstably ignites in the presence of H. The
\Hyd\ is then consumed by the rp-process, a sequence of rapid proton
captures onto seeds provided by the \He\ burning
\citep{wallace81:_explos,van94:_react,schatz98}.  Reaction network
calculations, both for single-zone calculations of unstable burning
\citep{koike99,schatz.aprahamian.ea:endpoint} and for steady burning
\citep{schatz99}, find that all of the hydrogen is consumed and that
the reactive flow reaches nuclei much heavier than \Fe.  In a recent
calculation, \citet{schatz.aprahamian.ea:endpoint} determined that the
rp-process ends in a closed SnSbTe cycle; the resultant ash
composition has a mean nuclear charge $\Zbar = 37$ and a mean nuclear
mass $\Abar = 79$.  Following the burst, the proton-rich elements 
quickly $\beta$-decay to more stable species such as \Ru.

The ratio of \Hyd\ to \He\ is not well determined.  When the
temperature in the crust exceeds roughly $10^{8}\usp\K$, the \Hyd\ in
the outer envelope is consumed on a timescale $\approx
6\ee{4}\usp\second\usp (X_\mathrm{CNO}/0.02)$ by the hot CNO cycle,
$X_\mathrm{CNO}$ being the mass fraction of CNO nuclei.  Hydrogen is
also consumed (in a thermally stable fashion) when the accretion rate
exceeds about $2\times 10^{-10}\usp\Msun\usp\yr^{-1}$.  During the
outburst decay, a significant fraction of the accumulated \Hyd\ can be
converted stably to \He, even after the last type I X-ray burst has
occurred.  We include this possibility in our calculations
(\S~\ref{sec:example}).  We note, however, that the photosphere would
still tend to be pure \Hyd, as $\Te$ is too cold for the \Hyd\ to be
consumed on a timescale over which differential sedimentation removes
CNO nuclei from the photosphere ($\sim 10\usp\second$;
\citealt{bildsten92}).  As a result, spectral fitting of the quiescent
flux will find that the ratio of radius to distance remains constant
even if $\Te$ varies.  Changes in the photospheric abundances, as
implied by, e.g., spectral features, more likely indicate quiescent
accretion.

\section{EQUATION OF STATE, SEDIMENTATION, AND THERMAL TRANSPORT}
\label{sec:microphysics}

The microphysics of the envelope enters equation~(\ref{eq:thermal})
explicitly through the opacity $\kappa$ and implicitly through
$\rho(y=p/g,T)$.  Before discussing the thermal transport and its
effect on the thermal structure of the envelope, we first review the
different physical regimes of the neutron star's envelope, the
equation of state, and the validity of treating the envelope as being
composed of distinct layers.  Figure~\ref{fig:phase} shows different
regimes of the $\rho$--$T$ plane for an envelope composed of a pure
\Hyd\ layer, of column depth $y = 10^{8}\usp\GramPerSc$,
superincumbent on a \Ru\ layer.  The top panel illustrates conditions
in the \Hyd\ layer; the bottom panel does likewise for the \Ru\ layer. 
In both panels, the circles denote the thermal structure found by
solving equation (\ref{eq:thermal}) for $\Te=2.1\ee{6}\usp\K$ (as
inferred for \aql; \emph{top curve}) and $\Te=8.8\ee{5}\usp\K$ (as
inferred for \cen; \emph{bottom curve}).  We discuss the calculations
for these sources in \S~\ref{sec:variation}.

\subsection{Equation of State}
\label{subsec:EOS}

\begin{figure*}[tbp]
    \centering\includegraphics{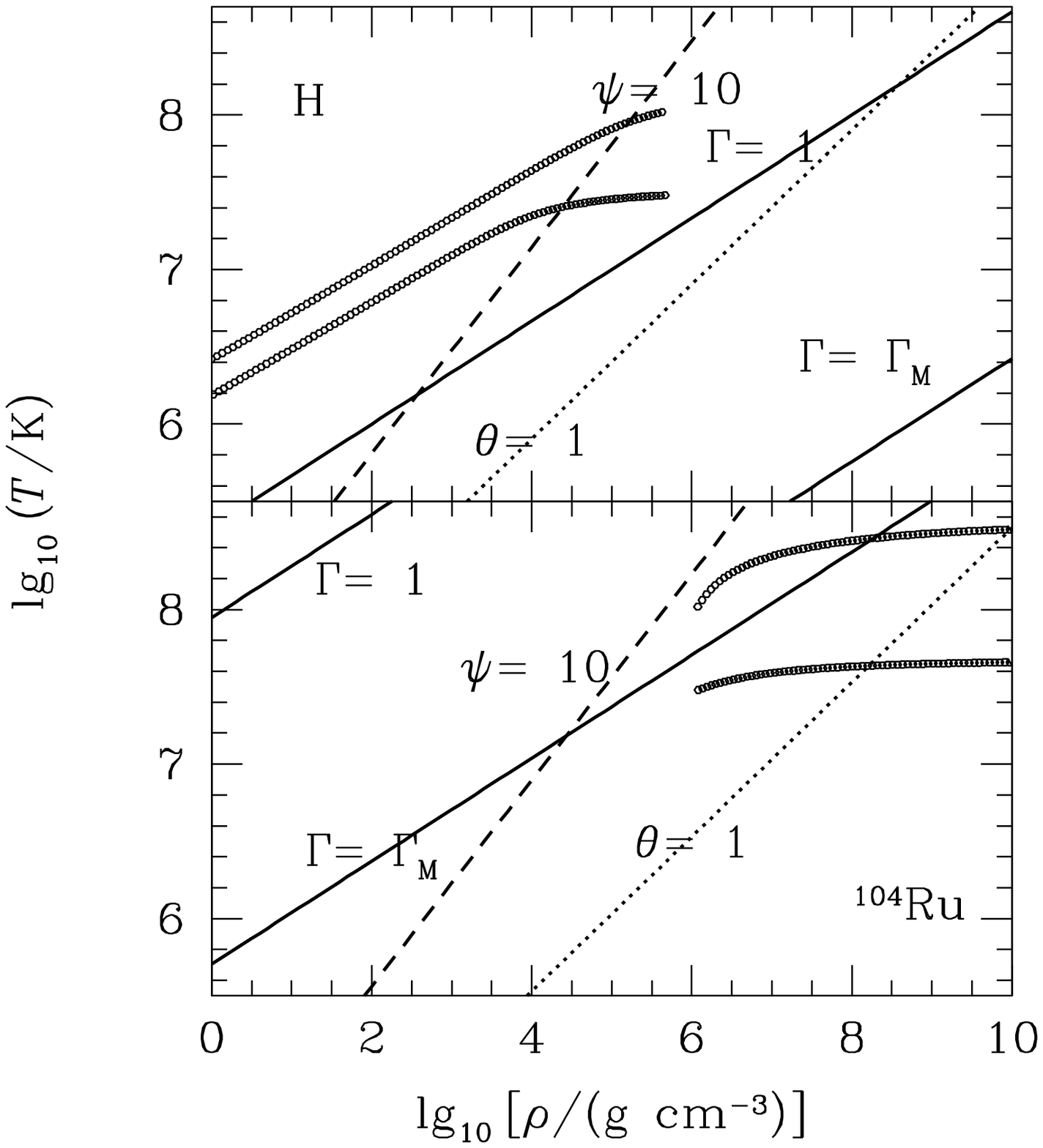}
    \caption{Schematic of different physical regimes for a quiescent
    neutron star envelope.  We show the conditions $\Gamma = 1$ and
    $\Gamma = \GammaM$ (\emph{solid lines}), $\theta = 1$ (\emph{dotted
    lines}) and $\psi = 10$ (\emph{dashed lines}), for both \Hyd\
    (\emph{top panel}) and \Ru\ (\emph{bottom panel}).  Circles denote
    the envelope structure for a two-layer (\Hyd\ over \Ru) envelope
    at $\Te = 2.1\ee{6}\usp\K$ (\emph{top curve}) and $\Te =
    8.8\ee{5}\usp\K$ (\emph{bottom curve}), as appropriate for \aql\ and
    \cen, respectively (see text).} \label{fig:phase}
\end{figure*}

The envelope of a neutron star is composed of electrons and ions. 
Electrostatic interactions between electrons are negligible throughout
most of the envelope for the temperatures of interest
\citep[see][]{chabrier98}, and the electrons are an ideal degenerate
Fermi gas for $\psi = \mu_\mathrm{e}/\kB T\gg 1$.  Here
$\mu_\mathrm{e}$ is the electron chemical potential, not including the
rest mass.  For $\psi \gg 1$, $\mu_\mathrm{e}\approx (\EF-\me c^{2})$,
where $\EF = (\me^{2}c^{4} + \pF^{2}c^{2})^{1/2}$ is the electronic
Fermi energy and $\pF = (3\pi^{2}\nel)^{1/3}\hslash$ is the Fermi
momentum.  We write the electron density $\nel \equiv \Ye\rho/\mb$
where \mb\ is the mean nucleon mass and delimit on
Figure~\ref{fig:phase} where the electrons are degenerate with the
condition $\psi = 10$ (\emph{dashed line}).  The electrons are
relativistic where $\pF/\me c \approx
[\Ye\rho/(10^{6}\usp\GramPerCc)]^{1/3} > 1$.

Where the envelope is composed of rp-process ashes, the total number
of species (each of density $n_{j}$) is likely quite large.
Electrostatic correlations between ions are parameterized by
\begin{eqnarray}
    \Gamma_{j} & = & \frac{Z_{j}^{2}e^{2}}{a_j\kB T}
    \label{eq:Gamma}  \\
     & \approx & Z_{j}^{5/3}
     \left(\frac{\Ye\rho}{10^{8}\usp\GramPerCc}\right)^{1/3} 
     \left(\frac{10^{8}\usp\K}{T}\right), \nonumber
\end{eqnarray}
$a_{j} = (3 Z_{j}/4\pi \nel)^{1/3}$ being the ion sphere radius for
species $j$.  For $\Gamma > \GammaM$, the plasma is a solid; $\GammaM$
is computed by equating the free energies of the liquid and solid phases
(Fig.~\ref{fig:phase}, \emph{solid line}).  We compute the ionic free
energy for the liquid phase, $1\le\Gamma < \GammaM$, from the fit of
\citet{chabrier98} and for the solid phase from the fit of
\citet{farouki93}.  In a recent calculation,
\citet{potekhin.chabirer:eos_solid} determined that $\GammaM = 175.0\pm
0.4$, with a further relative uncertainty of $\sim 10\usp\%$ arising
from electron screening.  The calculations in this paper are
insensitive, fortunately, to the precise value of $\GammaM$.

The ions are classical for
\begin{eqnarray}
      \theta & = & \frac{T}{T_\mathrm{p,ion}} = \frac{\kB T}{\hslash}
      \left(\frac{\Abar^{2}\mb^2}{4\pi\rho\Zbar^{2}e^2} \right)^{1/2}
      \nonumber \\
     & \approx & \left(\frac{T}{4.0\ee{7}\usp\K}\right)
     \left(\frac{10^{8}\usp\GramPerCc}{\rho}\right)^{1/2}
     \left(\frac{\Abar}{2\Zbar}\right) \gg 1,
  \label{eq:theta}
\end{eqnarray}
where $\theta$ is the temperature in units of the ion plasma
temperature.  The ratio of melting to plasma temperatures is
\begin{equation}
    \frac{T_\mathrm{M}}{T_\mathrm{p,ion}} = 3.4
    \left(\frac{175}{\GammaM}\right)
    \left(\frac{\Zbar}{30}\right)^{5/3}
    \left(\frac{2\Abar}{\Zbar}\right)^{2/3} 
    \left(\frac{10^{8}\usp\GramPerCc}{\rho}\right)^{1/6}
    \label{eq:ratio}
\end{equation}
and so $T\gtrsim T_\mathrm{p,ion}$ wherever the ions are composed of
high-$Z$ species and are in a liquid phase.  When $\theta < 1$
(\emph{dotted lines}, Fig.~\ref{fig:phase}), quantum lattice effects
become important.  As evident from Fig.~\ref{fig:phase}, in a \Hyd\
layer the ions are typically weakly coupled throughout, although one
should be careful about quantum plasma effects \citep[for a discussion,
see][]{potekhin97}.  In contrast, the ions in a \Ru\ layer are strongly
coupled wherever the electrons are degenerate.  Quantization of phonon
modes is not important, however, except in the very degenerate layers
where the thermal gradient is nearly isothermal.

\subsection{Sedimentation}\label{subsec:sedimentation}

The calculation in this paper presumes that the envelope is segregated
into layers.  An order-of-magnitude calculation illustrates the
timescale for the envelope to become stratified.  In a frame co-moving
with the mean ion center-of-mass (CM; here at rest), the continuity
equation for species $i$ is
\begin{equation}
  \frac{\dif n_i}{\dif t} = \grad \bvec{\cdot} \left(\D\grad n_i -
  n_i\vdr_{i}\right),
\label{eq:continuity}
\end{equation}
where $\D$ is the interspecies diffusion coefficient and $\vdr_{i}$ and
$n_{i}$ are respectively the drift velocity, relative to the mean ion
CM, and the number density of species $i$.  This equation neglects terms
arising from thermal diffusion, which are generally small in dense ionic
plasmas
\citep{stevenson.salpeter:phase,paquette.pelletier.ea:diffusion}.

The diffusion coefficient, $\D$, and the drift velocity, \vdr, are
both local quantities, i.e., they do not depend on $\grad \rho$ or
$\grad T$.  For a trace component (species ``2'') in a background
(species ``1'') these quantities are effectively fixed, and are
therefore related through Einstein's relation \citep[see,
e.g.,][]{landau.lifshitz:fluid}, $\vdr_{2} = \left[\D/(\kB T)\right]
\bvec{F}_2$ where $\bvec{F}_2$ is the force on the trace ion and the
combination $\D/(\kB T)$ is the mobility.  This relation holds where
the temperature scale height $-[\dif(\ln T)/\dif r]^{-1} \gg \Hp =
p/\rho g$, the pressure scale height.  Since $-\Hp[\dif(\ln T)/\dif r]
= \dif(\ln T)/\dif(\ln p) < (\partial\ln T/\partial\ln p)_{s} < 1$,
this requirement is trivially satisfied.  The force $\bvec{F}_{2}$ is
the sum of gravity and the electric field needed to keep the ions from
settling relative to the electrons \citep[see,
e.g.,][]{spitzer62:_physic,hameury.ea:settling},
\begin{eqnarray}
\bvec{F}_2	&=& A_2 \mb \bvec{g} - Z_2 e \bvec{E}\nonumber\\
		&=& \cases{%
	\left(A_2 - \frac {Z_2 A_1}{Z_1 + 1}\right) \mb \bvec{g},
		&non-degenerate\cr  
	\left(A_2 - \frac {Z_2 A_1}{Z_1}\right)\mb \bvec{g},
		&degenerate.}
\label{eq:F2}
\end{eqnarray}
The electric field is $e\bvec{E} = \mb \bvec{g} \left[A_1/(Z_1 +
1)\right]$ where the electrons are nondegenerate and $e\bvec{E} = \mb
\bvec{g} (A_1/Z_1)$ where they are degenerate and contribute most of the
gas pressure.

From continuity (eq.~[\ref{eq:continuity}]) the timescale for the trace
ions to move a distance $s$ is $\taudr =
\min(s^2\D^{-1},s\|\vdr_{2}\|^{-1})$, where the first term is the
diffusion timescale and the second is the drift timescale.  These
timescales are equal for $s = \Htr = \kB T/\|\bvec{F}_2\|$, which is
just the scale height for the trace ions.  For mass/charge ratios much
greater than unity and where the electrons are degenerate, $\Htr\ll\Hp$;
therefore over macroscopic scales $s\sim\Hp$, the relevant timescale is
the drift timescale $\taudr = s\|\vdr_{2}\|^{-1} = s F_{2}(\kB T/\D)$.
We can evaluate $\taudr$ once the mobility is determined.  For the
situation in question ($y \sim 10^8 \usp\GramPerSc$, $T \sim
10^8\usp\K$), the background matter exists in a liquid phase ($1 <
\Gamma_1 < \Gamma_{1,\mathrm{M}}$) (see Figure~\ref{fig:phase}).  We
follow the calculation of \citet{bildsten.hall:diffusion} and estimate
$\D$ from the Stokes-Einstein relation between the mobility and the drag
coefficient (for a liquid sphere), $\D/\kB T = 1/(4 \pi a_2 \eta)$.  In
this expression, $a_2$ is the radius of a charge-neutral (containing
$Z_2$ electrons) sphere around the trace particle and $\eta$ is the
viscosity of the material.  Fits to numerical simulations of
one-component plasmas (OCP) in the liquid regime find that the viscosity
is $\eta\approx (0.1\usp\gram\usp\cm^{-1}\usp\second^{-1})
(\rho/\GramPerCc) (\omega_\mathrm{p,1}/\second^{-1}) (a_1^{2}/\cm^2)
(\Gamma_1/10)^{0.3}$, where $a_1$ is the ionic spacing of the background
fluid and $\omega_\mathrm{p,1} = \left[4\pi \rho
(Z_1e/A_1\mb)^2\right]^{1/2}$ is the plasma frequency
\citep{donko.nyiri:viscosity,bildsten.hall:diffusion}.

Using the non-relativistic degenerate electron equation of state to
relate $p$ and $\rho$ and evaulating $\bvec{F}_{2}$ from
equation~(\ref{eq:F2}), we find from the Stokes-Einstein relation the
mobility,
\begin{equation}\label{eq:mobility}
   \frac{\D}{\kB T} \approx 1.2\ee{7}\usp\second\usp\gram^{-1}\usp
	\frac{A_1^{0.1} T_7^{0.3}}{Z_1^{1.3} Z_2^{0.3} \rho_5^{0.6}}.
\end{equation}
For brevity we scale the surface gravity, temperature, and density to $g
= g_{14}10^{14}\usp\cm\usp\second^{-2}$, $T = T_{7}10^{7}\usp\K$, and
$\rho = \rho_{5}10^{5}\usp\GramPerCc$.  The value $\D$ computed from
equation~(\ref{eq:mobility}) is comparable (within a factor of a few) to
that calculated by \citet{tanaka.ichimaru.87} for a strongly coupled OCP
under the conditions of interest.  It is somewhat larger than the value
estimated with the formalism of \citet{chapman.cowling} or
\citet{burgers69:composite_gases}, both of which are valid for weakly
coupled plasmas, $\Gamma\ll 1$ \citep[also
see][]{fontaine.michaud.79,paquette.pelletier.ea:diffusion}.

Upon substituting equation~(\ref{eq:mobility}) into the expression for
$\vdr_{2}$ and using \Hp\ as a fiducial lengthscale, one arrives at the
stratification timescale,
\begin{equation}\label{eq:tauD-number}
   \taudr \approx 10^5\usp\second 
	\left[\frac{Z_1^{3.9} Z_2^{0.3}\rho_5^{1.3}}{A_1^{1.8} g_{14}^2
	T_7^{0.3} \left(A_2Z_1 - A_1Z_2\right)}\right]. 
\end{equation}
For \Ru\ in \He\ at $\rho_{5} = 1$, $T_{7} = 10$, and $g_{14} = 1$
(appropriate for \aql; \S~\ref{sec:variation}),
equation~(\ref{eq:tauD-number}) implies that the \Ru\ ions settle in a
time of roughly 2\usp\hour.  For \Fe\ in \He, the timescale is roughly
7\usp\hour.  For the less dense regions of the envelope, the diffusive
time scale is of order seconds to minutes, so in the absence of any
circulation the envelope quickly stratifies after the end of the
accretion outburst.  Our calculations throughout the remainder of the
paper assume a fully stratified envelope.  We note that in diffusive
equilibrium, the boundary between the layers has a thickness $\sim
\Htr\ll\Hp$, which justifies our approximating the interface as a planar
surface.

\subsection{Thermal Transport}\label{subsec:thermal-transport}

For the temperatures and densities in the quiescent neutron star
envelope, the relevant opacities are Thomson scattering and free-free
absorption.  Because the outermost layers are composed of H and \He,
the ions are fully ionized throughout.  At typical envelope
temperatures, the dominant opacity is from free-free absorption,
\begin{equation}
    \opac_\mathrm{ff} \propto \nel T^{-7/2} \sum_{\jmath}
    Z_{j}^{2}Y_{j}g_{\mathrm{ff},j}
    \label{eq:free-free}
\end{equation}
where the Gaunt factor for species $j$, $g_{\mathrm{ff},j}$, contains
corrections for electronic Coulomb wavefunctions, degeneracy, and
relativistic effects.  There are no fits of $g_\mathrm{ff}$ that cover
the entire $(\rho,T)$ range appropriate for this problem.  The relation
between \Tb\ and \Te\ is most sensitive to \opac\ where the electrons
are semi-degenerate, however, so we use the fit from \citet{schatz99}
that is tuned to be accurate for $\psi\lesssim 10$ and moderately strong
Coulomb corrections, parameterized by $-4 < \ln\gamma^{2} = Z^{2}e^{4}
\me/(2\hslash^{2}\kB T) < 2$.  This fit is reasonably accurate
(fractional errors $\sim 10\usp\%$) when compared against the
calculations of \citet{itoh85:_relat,itoh91:_rossel_gaunt}.  We
calculate the Thomson scattering opacity by using a fit
\citep{buchler.yueh:compton} that reproduces the non-degenerate limit
\citep{sampson:opacity} and includes corrections for the relativistic
and degenerate electronic EOS.

As $\psi$ increases, the degenerate electrons become more efficient than
photons at transporting heat.  The electron thermal conductivity is
given in the relaxation-time approximation by the Wiedemann-Franz law,
\begin{equation}
    K = \frac{\pi^{2}}{3}\frac{\nel\kB^{2}T}{\me^{\star}}\tau,
    \label{eq:K}
\end{equation}
where $\me^{\star} = \EF/c^{2}$ is the effective electron mass, and
$\tau$ is the electron thermal distribution relaxation time.  In this
approximation, the relaxation time is the reciprocal sum over
electron-electron and electron-ion scattering relaxation times,
$\tau^{-1} = \tau_\mathrm{e,e}^{-1} + \tau_\mathrm{e,ion}^{-1}$.  In the
heavy element layer, the large values of $\Gamma$ (see
Fig.~\ref{fig:phase}) may inhibit stratification, and therefore we must
consider a multi-species plasma.  Where $\theta \gtrsim 1$, a reasonable
approximation (motivated by the additivity rule in multi-ionic EOS;
\citealt{potekhin99:_trans}) is to sum over inverse relaxation times for
each species, $\tau_\mathrm{e,ion}^{-1} = \sum_{j}
\tau_{\mathrm{e},j}^{-1}$, where $\tau_{\mathrm{e},j}$ are the separate
inverse relaxation times for electron-electron and electron-ion (from
species $j$) scattering, respectively.  We calculate the
electron-electron scattering relaxation time from the formalism of
\citet{urpin80:_therm}, as fit by \cite{potekhin97}.  The inverse
electron-ion scattering relaxation time is
\begin{equation}
    \tau_{\mathrm{e,ion}}^{-1} = \frac{4\pi e^{4}}{\pF^{2}v_\mathrm{F}}
    \frac{\rho}{\mb} \sum_{\jmath}Z_{j}^{2}Y_{j}\Lej.  \label{eq:tau}
\end{equation}
Here $v_\mathrm{F}=\pF/\me^{\star}$ is the electron velocity evaluated
at the Fermi surface, and $\Lej$ is the dimensionless Coulomb
logarithmic term that originates in the integration of the scattering
rate over electron phase space.  To evaluate $\Lej$, we use the fitting
formula of \citet{potekhin99:_trans}, which is straightforward to
implement for arbitrary $(Z,A)$.

Where $\theta\lesssim 1$, phonon modes begin to ``freeze out,'' and the
additivity rule (eq.~[\ref{eq:tau}]) becomes suspect.  In practice this
is not typically a concern, because scattering from charge
\emph{fluctuations} (impurity scattering) becomes more important than
electron-phonon scattering.  If the impurities are randomly distributed,
then the ``structure factor'' in the integration of the scattering
integral can be set to unity (Potekhin, private communication; also see
\citealt{itoh93}), and $\Leimp$ resembles that of the liquid
($\Gamma\ll\GammaM$) phase with a relaxation time depending on the rms
charge difference,
\begin{equation}
    \tau_\mathrm{imp}^{-1} = \frac{4\pi e^{4}}{\pF^{2}v_\mathrm{F}}
    \frac{\rho}{\mb}
    \sum_{\jmath}\left(Z_j-\Zbar\right)^{2}Y_{j}\Leimp.
    \label{eq:tau-imp}
\end{equation}
Thus the scattering differs from that in a liquid by a factor
\begin{equation}
    \frac{\ZZbar}{\Zbar^{2}} - 1 \equiv \frac{Q}{\Zbar^{2}}
    \label{eq:Q}
\end{equation}
With the structure factor in eq.~(8) of \citet{potekhin99:_trans} set to
unity, we find that the resulting $\Leimp$ is comparable to the fit
given by \citet{itoh93}.

How large might $Q$ be?  The output from a one-zone X-ray burst
nucleosynthesis calculation \citep{schatz.aprahamian.ea:endpoint} has
$Q/\Zbar^{2} = 233/37^{2} = 0.17$.  The computation of $\Lej$ assumes
that the separate species are arranged in a lattice; it is difficult to
imagine how this could come about in the case of an accreted neutron
star crust.  For the conditions of interest in this paper ($\rho
\lesssim 10^{10}\usp\GramPerCc$, $T\lesssim 10^{8}\usp\K$, and an
rp-process ash composition), $\tau_\mathrm{imp}^{-1} \lesssim \sum_{j}
\tau_{\mathrm{e},j}^{-1}$.  Impurity scattering is therefore not
dominant, unlike the case in the deep crust
\citep{brown:nuclear,gnedin.ea:thermal_relaxation}; it is also not
negligible, however, so the question arises as to how the two scattering
processes should add.  Such a calculation, while clearly important, is
beyond the scope of this paper.  We instead take a pragmatic approach
and use two different prescriptions:
\begin{equation}
   \tau_{e,\mathrm{ion}}^{-1} = \frac{4\pi e^{4}}{\pF^{2}v_\mathrm{F}}
   \frac{\rho}{\mb}
   \max\left(\sum_{\jmath}Z_{j}^{2}Y_{j}\Lej,\frac{Q}{\Abar}\Leimp\right)
   \label{eq:tau-eff}
\end{equation}
and
\begin{equation}
   \tau_{e,\mathrm{ion}}^{-1} = \frac{4\pi e^{4}}{\pF^{2}v_\mathrm{F}}
   \frac{\rho}{\Abar\mb} \left(\Zbar^2 \bar{\Lambda}_{e,\mathrm{ion}} +
   Q\Leimp \right).  \label{eq:tau-mi}
\end{equation}
Here $\bar{\Lambda}_{e,\mathrm{ion}}$ is the Coulomb logarithm for a
single ion of charge number $\Zbar$ and mass number $\Abar$.  Both
approaches give comparable results in the Debye screening limit
($\theta\to 0$) and in the limit ($\theta \gg 1$) where impurity
scattering dominates.  In the regime where both impurity and phonon
scattering are comparable, the second prescription
(eq.~[\ref{eq:tau-mi}]) gives a larger $\tau_{e,\mathrm{ion}}^{-1}$
and hence a smaller $K$.  For purposes of comparison
(\S~\ref{sec:example}) we compare the conductivity of a pure state,
e.g., \Ru\ with that obtained using equation~(\ref{eq:tau-mi}); this
gives the largest variation in $K$.

\subsection{Sensitivity}
\label{subsec:sensitivity}

The choice of input physics can dramatically affect the relation $\Te
= \Te(\Tb)$.  The greatest uncertainty lies with the calculation of
conductive opacities around the melting point, $\Gamma \approx
\GammaM$, and in the crystalline phase.  This is partly due to our
ignorance of the exact composition of the envelope.  We consider in
\S\S~\ref{sec:example} and \ref{sec:variation} different possibilities
for the composition of the heavy elements: \Fe, \Ru, and the
rp-process ashes.  In this section we consider how a different
prescription for the electron-ion conductivity
\citep{flowers81,itoh83,itoh93} than our adopted formulae
\citep{potekhin99:_trans} would change our results.  By looking at the
sensitivities of our results to the choice of conductivity, we can
understand how our results vary in response to the general
uncertainties in input physics.

\begin{figure*}[tbp]
    \centering\includegraphics[width=\textwidth]{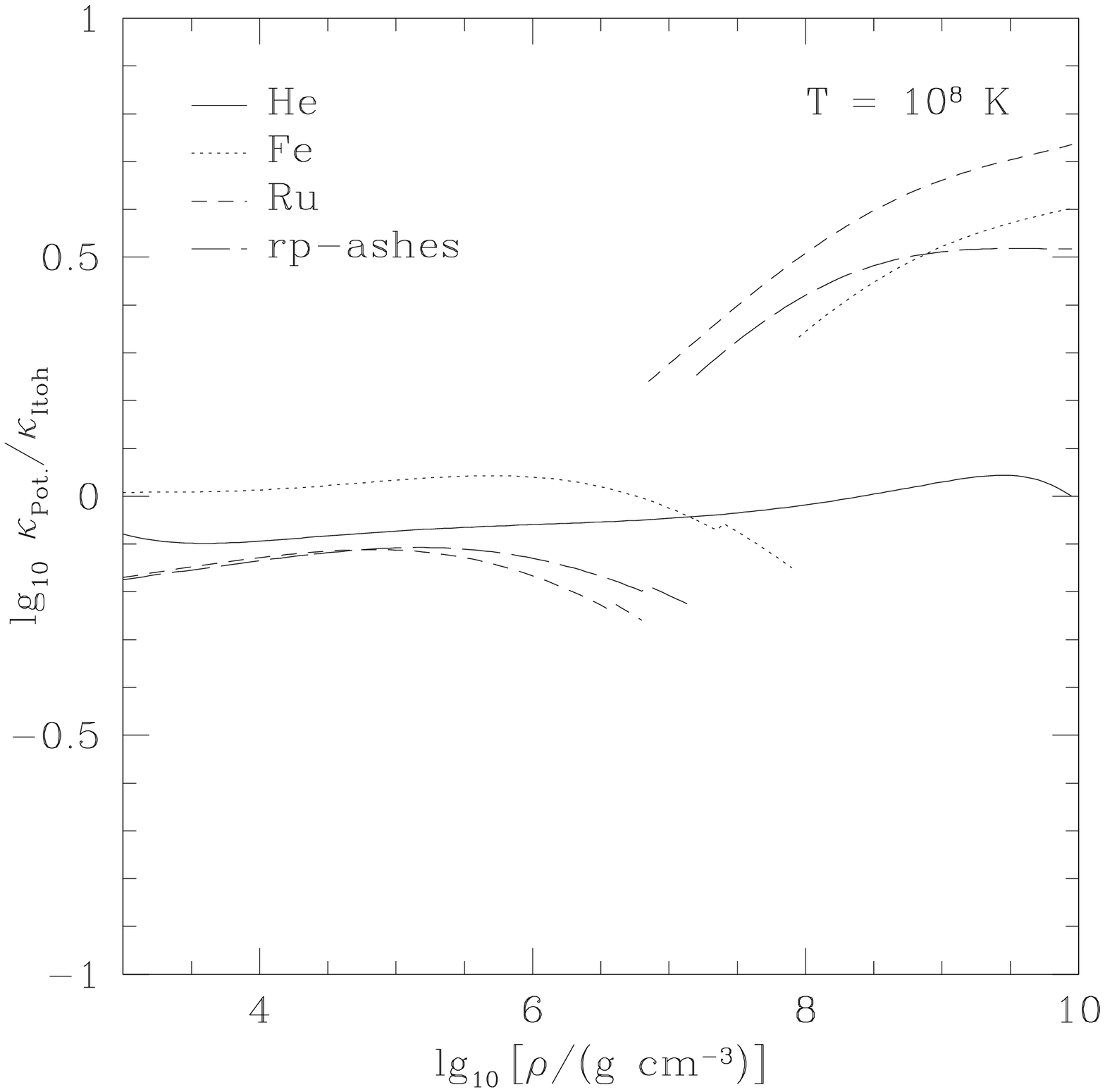} 
    \caption{Ratio of the conductive opacity as fitted by
    \protect\citet{potekhin99:_trans}, $\kappa_\mathrm{Pot.}$, to that
    fitted by \protect\citet{itoh83}, $\kappa_\mathrm{Itoh}$, for
    various plasma compositions as a function of density at a
    temperature $T = 10^8 \usp\K$.  The discontinuities occur at the
    liquid-solid phase transition.  They reflect our choice to use (for
    the results of \protect\citealt{itoh83}) the liquid metal
    approximation for $\Gamma < \GammaM$ and the electron-phonon
    scattering for $\Gamma \ge \GammaM$.}
    \label{fig:cond_opac_comp}
\end{figure*}

Figure \ref{fig:cond_opac_comp} highlights this problem.  Away from the
melting point in the liquid regime, the fitting formulae given by
\citet{potekhin99:_trans}, $\kappa_\mathrm{Pot.}$, and \citet{itoh93},
$\kappa_\mathrm{Itoh}$, are in good agreement,
$|1-\kappa_\mathrm{Pot.}/\kappa_\mathrm{Itoh}| \approx
5\usp\%\mbox{--}60\usp\%$.  The agreement begins to fall apart near the
melting point and the conductivities in the crystalline regime differ by
a factor of 2--3 \citep[for a discussion, see][]{potekhin99:_trans}.
Because our conductivity is uncertain, in any case, for a multi-species
plasma (\S~\ref{subsec:thermal-transport}), we consider how variations
in the thermal conductivity affect the relation between $\Tb$ and $\Te$.

For a neutron star envelope of fixed composition with a given \Te, the
temperature profile $T(y)$ can roughly be divided into three regions:
the radiative zone, $\opaccond \gg \opacrad$; the sensitivity strip,
$\opaccond \approx \opacrad$; and the isothermal zone $\opaccond \gg
\opacrad$.  The sensitivity strip is so named because changes to the
conductive opacity in this region strongly affect the temperature
profile of the envelope
\citep{gudmundsson83,potekhin97,ventura.potekhin:neutron}.  A change in
the conductive opacity in the sensitivity strip changes the region where
the sensitivity strip lies.  Since this region controls the transition
from a power law radiative solution to a isothermal zone solution, it is
critical that the conductive opacity be well-understood here.

The location of the sensitivity strip \yss\ is roughly where $\opaccond
\approx \opacrad$.  Setting $\opaccond = C \opacrad$ (where $C$ is an
arbitrary constant that contains our uncertainty regarding the
composition and scattering integrals in $\opaccond$ and $\opacrad$),
using an ideal gas equation of state, and setting factors of order unity
to unity, we find that
\begin{equation}\label{eq:yss}
	\yss \approx 2.2\ee{3}\usp\GramPerSc \left[ C^{-1/3} 
	\frac{A^2 Z }{(Z +1) g_{14}} T_7^{17/6}\right].
\end{equation}
To relate \yss\ to \Te, we insert the solution to the flux equation
(eq.~[\ref{eq:thermal}]) in the radiative zone.  Since the dominant
opacity is from free-free absorption, we can take as our opacity $\opac
\propto \rho T^{-7/2}$.  Inserting this into equation~(\ref{eq:thermal})
and again using an ideal gas equation of state, we find that $y^2
\propto T^{17/2}$.  Solving for $T$ and inserting all the appropriate
numerical factors, we have
\begin{equation}\label{eq:T-rad-zone}
	T_7 = 0.16 y^{4/17} 
	\left[\frac{Z^3 g_{14} T_\mathrm{E,6}^4}{A(Z+1)}\right]^{2/17}.
\end{equation}
Inserting the expression for \yss, equation~(\ref{eq:yss}), into
equation~(\ref{eq:T-rad-zone}), one finds that the temperature in the
sensitivity strip scales as $T_\mathrm{ss} \propto C^{-4/17}$.  To
estimate how \Te\ scales with the microphysical input, we note that if
$T_\mathrm{ss} \approx \Tb$ then $\Te \propto C^{1/2}$.  Therefore, this
calculation is moderately sensitive to fractional uncertainties of order
$10\usp\%$ in the input physics.  Since the sensitivity strip is in the
regime where the envelope matter remains a liquid
\citep{ventura.potekhin:neutron}, our sensitivity to the microphysics of
the crystalline matter is relatively small, especially for the surface
temperatures ($\Te \sim 10^6\usp\K$) of interest.

\section{THE VARIATION OF EFFECTIVE TEMPERATURE}\label{sec:example}

Having laid out our microphysical tools, we are now ready to explore how
the changing envelope stratification varies the relation between the
deep crust temperature $\Tb$ and the effective temperature $\Te$.  To do
this, we adopt a two-layer model with a variable column depth $\yi$ of
the top layer.  The outer layer is composed of \Hyd\ or \He, and the
inner layer either pure \Fe, \Ru, or ashes from rp-process burning.  We
integrate equation~(\ref{eq:thermal}) numerically using an Adams
predictor-corrector method \citep{hindmarsh83:_odepac_ode}.  As a
boundary condition for equation~(\ref{eq:thermal}), we apply the
Eddington approximation\footnote{The value of $\Tb$ for a given $\Te$ is
insensitive to the precise location of the photosphere, so this
approximation is sufficient for our purposes.} at the photosphere,
$\opac(\rho,\Te) y_\mathrm{ph} = 2/3$.  For a given $\Te$, we then
integrate equation~(\ref{eq:thermal}) inwards to $\yb =
10^{14}\usp\GramPerSc$.  At this column, the thermal gradient becomes
nearly isothermal, and $\Tb = T(y=\yb)$ is approximately the interior
temperature.  The inverse relation $\Te(\Tb)$ is then found by
iteration.  As a check, we compared our calculations to those of
\citet{potekhin97} for a \Fe\ envelope and a ``fully accreted'' envelope
(H/He/C/Fe layers).  For a given \Tb, the fractional difference between
our value of \Te\ and that computed from the fitting formula of
\citet{potekhin97} is of order 5\usp\%, with the largest deviation
occurring when the \Hyd/\Fe\ interface is in the sensitivity strip.

\begin{figure*}[tbp]
    \centering\includegraphics{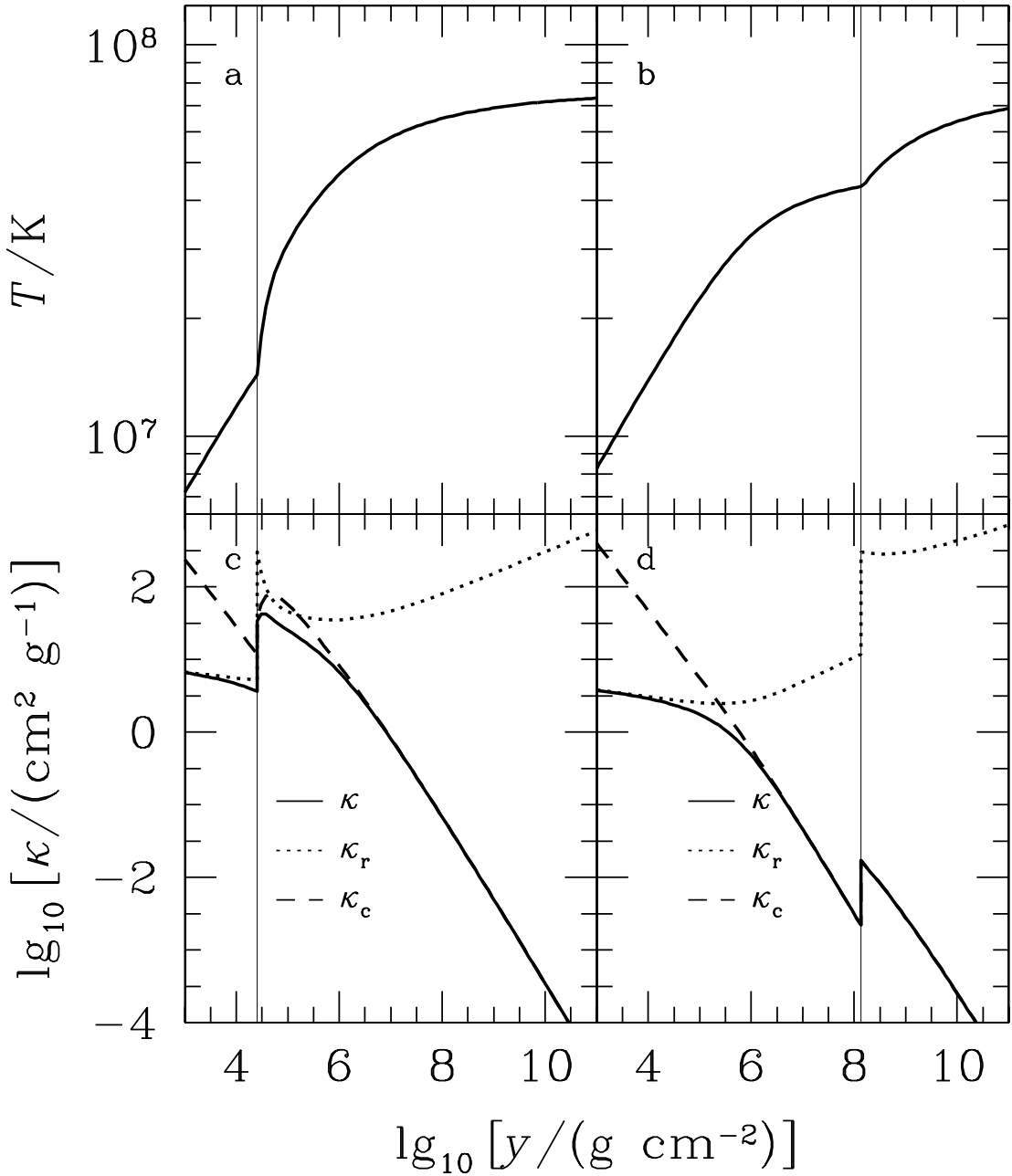}
    \caption{Temperature and opacities for a two-layer (\Hyd\ over
    \Ru) envelope.  The interface between the layers is denoted by
    the thin solid vertical lines: $\yi = 2.5\ee{4}\usp\GramPerSc$
    (\emph{left panels}) and $\yi = 1.3\ee{8}\usp\GramPerSc$
    (\emph{right panels}).  The top panels depict the temperature; the
    bottom panels depict the total opacity (\emph{solid lines}),
    radiative opacity (\emph{dotted lines}), and conductive opacity
    (\emph{dashed lines}).}
    \label{fig:opacity}
\end{figure*}

To illustrate how the opacity changes with the variation in the
\emph{location} of the interface, we show in Figure~\ref{fig:opacity} a
two-layer neutron star envelope (\Hyd\ superincumbent on \Ru) with
$\yi=10^{4.4}\usp\GramPerSc$ (panels a, c) or
$\yi=10^{8.13}\usp\GramPerSc$ (panels b, d).  In both cases $\Tb =
7.5\ee{7}\usp\K$.  The two top panels (a, b) depict the temperature,
while the bottom panels (c, d) display the total opacity (\emph{solid
line}), radiative opacity (\emph{dotted line}), and conductive opacity
(\emph{dashed line}).  When the interface is at a low column, both the
radiative and conductive opacities play a role.  At higher column, the
conductive opacity dominates at the location of the interface.  In both
cases there is a substantial increase in the opacity of a \Ru\ layer
from a \Hyd\ layer, reflecting the increase in the ion charge for
bremsstrahlung and electron-ion scattering.

At low densities, the opacity is dominated by radiative processes
(mostly free-free).  For a free-free dominated envelope, $T(y) \propto
y^{4/17}$ (eq.~[\ref{eq:T-rad-zone}]); as a result, along the trajectory
$\{y,T(y)\}$, the free-free opacity is $\opac_\mathrm{ff} \propto
y^{-1/17}$ (Fig.~\ref{fig:opacity}, \emph{dotted lines}) and is nearly
constant.  As the composition changes from \Hyd\ to \Ru, the opacity
jumps by a factor $\approx 44^{2}/104$.  Where the electrons are
degenerate and the heat transport set by electron conduction, $\dif
T/\dif y\to 0$, and the Gaunt factor scales as $\nel^{-2/3}$, so
$\opac_\mathrm{ff} \propto y^{1/4}$.  In the limit where the electrons
are degenerate and relativistic, the electron conductive opacity scales
as $\opaccond\propto [T(y)]^{2}/y$, and $T(y)$ is nearly constant.  The
jump at the interface is $\Delta\opaccond< Z^{2}/A=44^{2}/104$ because
the stronger ion-ion correlations (parameterized by $\Gamma$) decrease
the scattering rate $\tau_{\mathrm{e},j}^{-1}$ and offset the increase
from the larger ionic charge.

\begin{figure*}[tbp]
    \centering \includegraphics{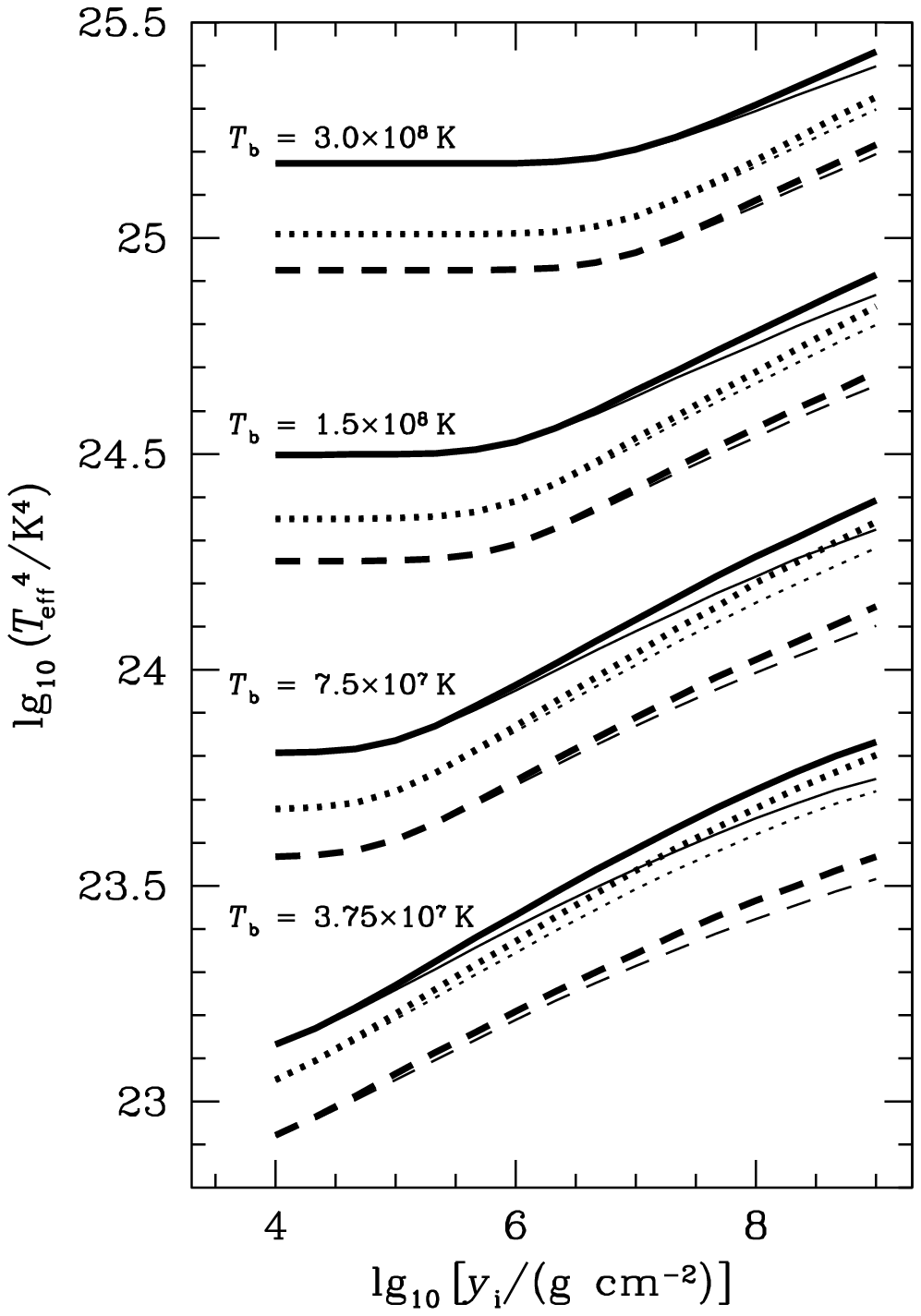}
    \caption{Variation of $\Te^{4}$ with interfacial column $\yi$ for
    a two-layer envelope with a fixed inner boundary temperature
    $\Tb$.  The top layer is either \Hyd\ (\emph{thin lines}) or \He\
    (\emph{thick lines}); the bottom layer is either \Fe\ (\emph{solid
    lines}), \Ru\ (\emph{dotted lines}), or a very impure ($Q = 233$;
    $K$ calculated in the mean-ion approximation) mixture from an
    rp-process burst (\emph{dashed lines}).  Each group of curves
    corresponds to $\Tb = 3.75\ee{7}\usp\K$, $7.5\ee{7}\usp\K$,
    $1.5\ee{8}\usp\K$, and $3.0\ee{8}\usp\K$.}
    \label{fig:vary}
\end{figure*}

Figure~\ref{fig:vary} shows the emergent flux ($\Te^4$) for a
two-layer neutron star envelope as a function of $\yi$.  The top layer
is either pure \Hyd\ (\emph{thin lines}) or \He\ (\emph{thick lines}),
and the bottom layer is composed of either \Fe\ (\emph{solid lines}),
\Ru\ (\emph{dotted lines}), or an rp-process mixture with the mean-ion
approximation (eq.~[\ref{eq:tau-mi}]; \emph{dashed lines}).  For each
composition, we used 4 bottom temperatures: $\Tb = 3.75\ee{7}\usp\K$,
$7.5\ee{7}\usp\K$, $1.5\ee{8}\usp\K$, and $3.0\ee{8}\usp\K$.  The
difference in $\Te^{4}$ between an outer \Hyd\ layer and an outer \He\
layer is typically smaller than the difference between the different
ash compositions and locations of the interface.  Were we to calculate
$\tau_\mathrm{e,ion}$ using equation~(\ref{eq:tau-eff}) rather than
equation~(\ref{eq:tau-mi}), the resulting curves would have lain
between those of an \Ru\ layer (\emph{dotted lines}) and the rp-ashes
with the mean-ion approximation (\emph{dashed lines}).  The emergent
flux is insensitive to the location of the interface for $y\lesssim
10^{5}\usp\GramPerSc$; as the interface moves deeper the lower opacity
of \Hyd\ layer reduces $|\dif T/\dif y|$ so that $\Te$ increases for a
fixed $\Tb$.  Note also that the profile for the multi-species ash has
a lower $\Te$ (higher opacity) than that of a pure \Ru\ layer despite
having a smaller $\Zbar$.  Because the $\Gamma_{j}$ for each species
in a multi-component mixture is reduced relative to that for a pure
species, $\Lej$ is larger at high densities, where $\theta \lesssim
1$.

By how much can the depth of the outermost layer vary?  The maximum
depth of the \Hyd\ layer is set by the reaction
$\nuclei{}{p}(e^{-},\nu)\nuclei{}{n}$.  This reaction occurs for $\EF
> \mn - \mpr + \me = 1.29\usp\MeV$, or $\rho\approx
1.3\ee{7}\usp\grampercc$.  With our assumed surface gravity, the
corresponding column is $y = 1.6\ee{10}\usp\GramPerSc$.  For a pure
\He\ layer, the maximum depth would be where the strongly screened
3$\alpha$ reaction \citep{fushiki87:_s} ignites: $\rho >
10^{8}\usp\grampercc$ for temperatures $<10^{8}\usp\K$.  Accretion to
this depth requires a very slow accretion rate over a long time.  For
the column accretion rate needed to power the quiescent thermal
emission, $\dot{m}\approx 1\usp\GramPerSc\usp\second^{-1}$, the time
needed to accrete \Hyd\ to the electron capture depth is $500\usp\yr$. 
This could possibly occur for long-recurrence time transients such as
\source{KS}{1731}{-260} \citep{rutledge.ea.01:ks1731,wijnands:ks1731}. 
For short recurrence-time transients, such as \aql, the \Hyd\ layer
cannot be appreciably thicker than where \He\ unstably ignites,
$y_\mathrm{ign}\approx 10^{8}\usp\GramPerSc$ \citep[see][and
references therein]{bildsten:thermonuclear}.  How \emph{thin} the
light-element layer might be is more difficult to determine.  As noted
in \S\ref{sec:how-it-works}, the column of \Hyd\ deposited after the
last burst for spherically symmetric accretion is $\lesssim
y_\mathrm{ign}$.  If the accretion were to occur onto only a small
fraction of the surface and then later spread over the surface, than
the residual column could be much less than $y_\mathrm{ign}$.

\section{AQL~X-1, CEN~X-4, AND SAX~J1808.4$\bvec{-}$3658}
\label{sec:variation}

\begin{figure*}[tbp]
    \centering \includegraphics{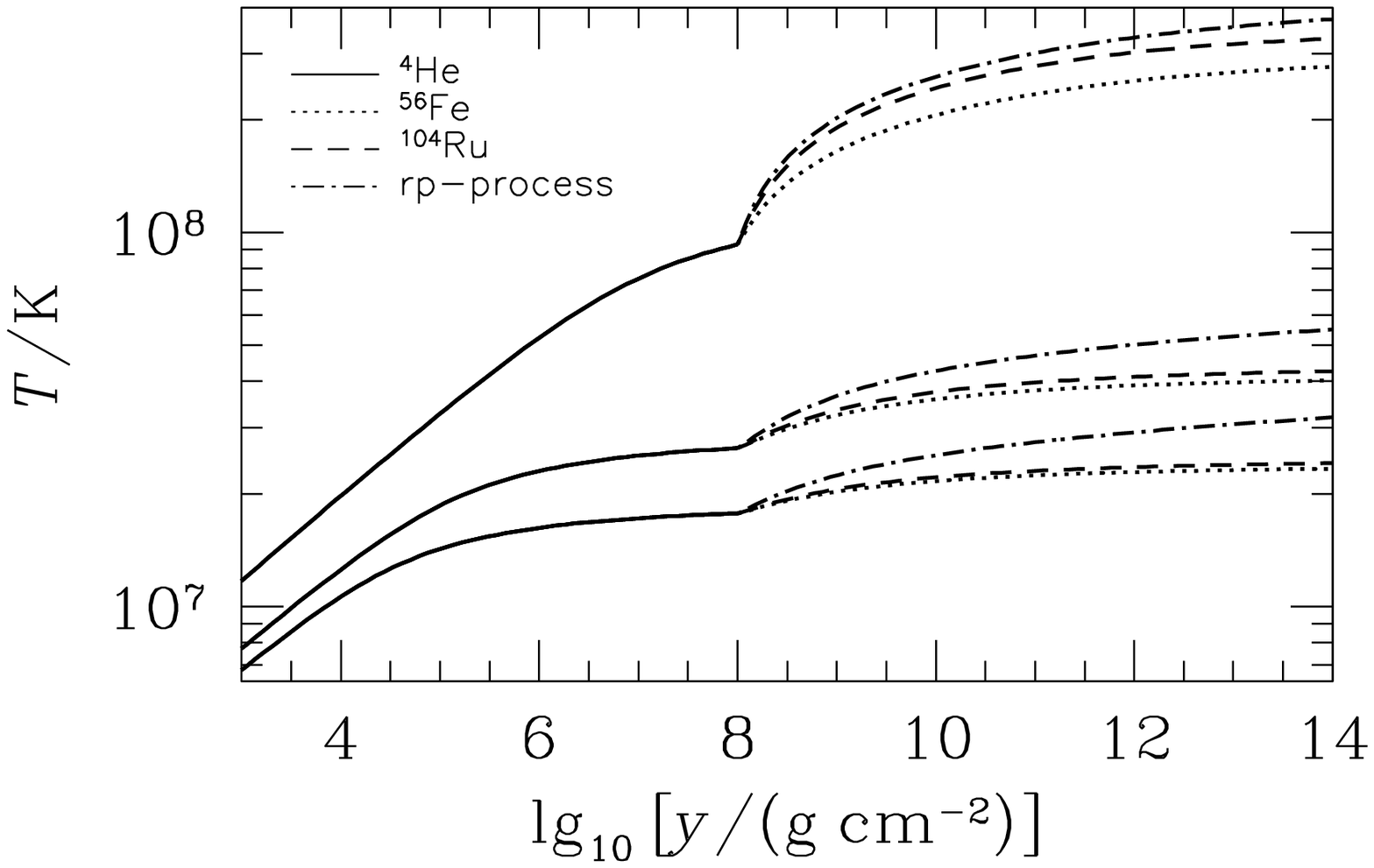}
    \caption{Thermal structure of a two-layer neutron star envelope
    with the interface between the layers at
    $\yi=10^{8}\usp\GramPerSc$.  The three structures, from top to
    bottom, are for $\Te = 2.1\ee{6}\usp\K$ (\aql), $\Te =
    8.8\ee{5}\usp\K$ (\cen), and $\Te = 6.8\ee{5}\usp\K$ (\sax).  The
    composition of the outer layer is \He\ (\emph{solid line}).  The
    composition of the inner layer is either \Fe\ (\emph{dotted
    line}), \Ru\ (\emph{dashed line}), or rp-process ashes with the
    mean-ion approximation (eq.~[\protect\ref{eq:tau-mi}]) for the
    conductivity (\emph{dot-dashed line}).}
    \label{fig:structure-aqlcen}
\end{figure*}

Having explored the variation induced in $\Te$ by varying $\yi$ and the
composition of the envelope, we now describe the thermal structure of
\aql, \cen, and \sax.  \aql\ and \cen, despite having similar binary
orbital periods (\unit{19}{hr} and \unit{15.1}{hr}, respectively), have
very different outburst morphologies: \aql\ goes into an $\approx
\unit{30}{d}$ outburst on a roughly yearly basis, while \cen\ has had
just two recorded outbursts (only one of which contributed significantly
to the total observed fluence) in the past \unit{33}{yr}.  \sax\ is
distinguished from both \cen\ and \aql\ by virtue of having pulsations
\citep{wijnands98} in the persistent emission; its orbital period is
also much shorter (\unit{2.01}{hr}; \citealt{chakrabarty98b}), and it
quite possibly accretes from a substellar mass companion (this also
explains its low time-averaged accretion rate, $\langle\dot{M}\rangle
\sim 10^{-11}\usp\Msun\usp\yr^{-1}$;
\citealt{bildsten.chakrabarty:bd_saxj1808}).
Figure~\ref{fig:structure-aqlcen} displays a summary of calculations for
these three objects: from top to bottom, \aql\ ($\Te =
2.1\ee{6}\usp\K$), \cen\ ($\Te = 8.8\ee{5}\usp\K$), and \sax\ ($\Te =
6.8\ee{5}\usp\K$).  We fix the composition of the outer layer to be \He\
(\emph{solid line}) with a column $\yi = 10^{8}\usp\GramPerSc$, and vary
the composition of the inner layer between \Fe\ (\emph{dotted line}),
\Ru\ (\emph{dashed line}), and rp-process ashes (\emph{dot-dashed
line}).  We now explain each calculation in more detail.

\chandra\ observations of \aql\ \citep{rutledge.ea.01b:aqlx1} find that
$\kB\Teo = 135^{+18}_{-12}\usp\eV$.  For a fiducial neutron star of mass
$M=1.4\usp\Msun$ and $R=10\usp\km$, the proper effective temperature is
$\Te = \Teo(1+z)=2.1\ee{6}\usp\K$.  At this temperature and $\yi$, the
fractional difference in $\Tb$ between an outer layer composed of \Hyd\
and one composed of \He\ is $\Tb(\Hyd)/\Tb(\He)-1 < 0.02$.  Using
Figure~\ref{fig:vary}, the variation in the emergent flux induced by a
changing $\yi$ is, at this $\Tb$, roughly a factor of 2; changing the
composition of the ash from \Fe\ to rp-process ashes increases $\Te^{4}$
by a factor $\approx 1.6$.

This calculation is consistent with widely spaced observations of \aql,
which find variability by a factor of 1.9 over a timescale of years
(i.e., over several outburst/recurrence timescales).  We cannot explain
changes on timescales less than this, however.  As mentioned in
\S~\ref{sec:introduction}, \citet{rutledge.bildsten.ea:intensity}
observed \aql\ with \chandra\ on four successive occasions following the
November 2000 outburst.  The intensity was observed to decrease by a
factor of $\approx 0.5$ over 3 months and then increase by a factor of
$\approx 1.4$ over 1 month.  In addition, short timescale ($<
10^4\usp\second$) variability was evident in the last observation.  Our
present calculation does not address the short timescale variability.
The fact that the intensity first decreased and then increased also
cannot be accomodated in the scenario we outline in this paper.
Moreover, the timescale for sedimentation (eq.~[\ref{eq:tauD-number}])
is far too brief.

For \cen, \chandra\ observations find that $\kB\Te = (76\pm7)\usp\eV$
\citep{rutledge.ea.01a:cen}.  The last known accretion outburst occurred
in 1979.  Assuming no outbursts have occurred since then, the crust has
had several thermal times to relax.  As with \aql, we again set the
outer layer to be pure \He\ (Fig.~\ref{fig:structure-aqlcen};
\emph{solid line}).  The reason we choose the outer layer to be \He\
reflects our prejudice that \Hyd\ is consumed as the accretion rate
decreases at the end of the outburst.  Changing the outer layer to \Hyd\
increases $\Tb$ by a factor $\Tb(\Hyd)/\Tb(\He)-1 < 0.08$.  For
transients with recurrence times of decades or longer, such as \cen\ and
\source{KS}{1731}{-260}, it is possible that residual accretion could
substantially \emph{increase} the depth of the \Hyd/\He\ layer.
\citet{wijnands:ks1731} and \citet{rutledge.ea.01:ks1731} found the
luminosity of \source{KS}{1731}{-260} to be $\approx
3\ee{33}\usp\ergspersecond$; this constrains the quiescent accretion
rate to $\dot{M}_{q} \lesssim 2\ee{-13}\usp\Msun\usp\yr^{-1}$.
Accretion at this limiting rate increases $\yi$ by
$10^{8}\usp\GramPerSc$ every 30 years.  By itself, this can change the
brightness by a factor of $1.2$ on this timescale.  This effect may be
dwarfed by the thermal relaxation of the crust, however, which also
occurs over a timescale of decades \citep{rutledge.ea.01:ks1731}.  A
simulation of the time-dependent thermal luminosity for such sources is
beyond the scope of this introductory paper; for now we just note this
interesting possibility.

Finally, we turn our attention to \sax.  This source is rather dim in
quiescence
\citep{stella00:quiescent_1808,dotani.asai.ea:decrease,wijnands02:_very_sax_j1808}.
We estimate the surface effective temperature from the flux reported by
\citet{wijnands02:_very_sax_j1808}; note that this is not a bolometric
flux, and may also include a contribution from a power-law.  For the
estimated temperature, the overall stratification is similar to that of
\cen, with $\Tb\approx 2.8\ee{7}\usp\K$.  The variation in $\Te$ is
approximately a factor of 4 at $\Te < 10^{6}\usp\K$.  Given its short
recurrence time and cold $\Te$, \sax\ may be the best source from which
to observe the effect described in this paper.  If the previous few
years are typical, then either \chandra\ or \xmm\ can likely observe
different quiescent epochs.

\section{SUMMARY AND DISCUSSION}\label{sec:summary}

In summary, we find that (1) the timescale for the neutron star envelope
to segregate into layers is much less than the outburst recurrence time,
making the surface effective temperature sensitive to the mass of
\Hyd/\He\ remaining on the surface at the end of an outburst; (2)
variations in the composition of the heat-blanketing envelope can lead
to variability, by a factor of 2 to 3, in the \emph{thermal} quiescent
flux from neutron star SXTs; (3) the crust temperatures of \aql, \cen,
and \sax, for the quoted effective temperatures, are respectively
$3.3\ee{8}\usp\K$, $4.9\ee{7}\usp\K$, and $2.8\ee{7}\usp\K$ with a
fractional uncertainty, from $\yi$ and the ash composition, of roughly
$20\usp\%$.

The measured $\Te$ and inferred $\Tb$ of \aql\ have interesting
implications for the interior temperature of the neutron star.  As
described in \citet{brown:nuclear}, when the temperature in the crust is
sufficiently hot, there is an inversion of $\dif T/\dif r$: heat flows
\emph{inward} from the crust to the core, where it is balanced by the
neutrino luminosity $L_{\nu}$.  A calculation similar to that described
in \citet{brown.ushomirsky:rmodes}, with the ``moderate superfluid''
case discussed in \citet{brown:nuclear} finds that the ratio of
quiescent photon luminosity to neutrino luminosity is
$L_\mathrm{q}/L_{\nu} = 3.6$.  Our choice of superfluid transition
temperatures in these calculations are similar to those employed by
\citet{yakovlev.kaminker.ea:neutrino} in fitting to observations of
cooling neutron stars.  This suggests that for sources with higher
$\langle L\rangle$ than \aql, the simple distance-independent relation
between the quiescent flux and the time-averaged outburst flux
\citep{brown98:transients} does not exactly hold.  For sources with
lower $\langle L\rangle$, neutrino cooling is not important, unless an
enhanced (beyond modified Urca; see \citealt{colpi.geppert.ea:charting})
neutrino emissivity operates.  A direction for future work would be to
incorporate recent fits \citep{kaminker.haensel.ea:nucleon} of cooling
neutron star observations.

We reiterate that the best place to observe the effect of a changing
light-element layer mass is a source such as \sax, which has both a low
$\Te$ \citep{wijnands02:_very_sax_j1808} and a short recurrence
timescale (2 outbursts in the past 6 years).  At lower $\Tb$, the
effective temperature can vary by as much as a factor of 4, while the
frequent outbursts allow comparison between different envelope layerings
for a fixed core temperature.

\acknowledgements

We thank A. Potekhin for educating us on the thermal conductivity and
for carefully reading the manuscript and F. Timmes for providing us with
tables and interpolation routines for the electronic equation of state.
This work is partially supported by the Department of Energy under grant
B341495 to the Center for Astrophysical Thermonuclear Flashes at the
University of Chicago, by the National Science Foundation under Grants
PHY 99-07949 and AST 01-96422, and by NASA through grant NAG 5-8658.
E. F. B.  acknowledges support from an Enrico Fermi fellowship.
L. B. is a Cottrell Scholar of the Research Corporation.


\clearpage


\clearpage

\clearpage

\clearpage

\clearpage

\clearpage

\end{document}